\documentclass[twocolumn,secnumarabic,amssymb, nobibnotes, aps, prd]{revtex4-2}
\usepackage{physics}
\usepackage{amsmath,amssymb,amsfonts,color,epsfig}
\usepackage{yfonts}
\usepackage{graphicx}
\usepackage{subfigure}
\usepackage{setspace}
\usepackage{physics}
\usepackage{mathtools}
\usepackage{bm}
\usepackage{tikz-cd}
\usepackage{tikz}
\usepackage{booktabs}
\usepackage{tabularx}
\usetikzlibrary{decorations.markings}
\usepackage{framed}
\usetikzlibrary{shapes.geometric, arrows, positioning}
\usepackage[T1]{fontenc}
\usepackage{hyperref}
\hypersetup{
    colorlinks=true,
    linkcolor=blue,
    citecolor=blue,
    urlcolor=blue
}

\newcommand{\nn}{\nonumber}

\begin{document}
%%%%%%%%%%%%%%%%%%%%%%%%%%%%%%%%%%%%%%%%%%%%%%%%%%%%%%%%%%%%%%%%%%%%%%%%%%%%
\title{Saving the Unruh Signal: Coherent Cancellation of Spontaneous Emission with Entangled Detectors}
%%%%%%%%%%%%%%%%%%%%%%%%%%%%%%%%%%%%%%%%%%%%%%%%%%%%%%%%%%%%%%%%%%%%%%%%%%%%

%%%%%%%%%%%%%%%%%%%%%%%%%%%%%%%%%%%%%%%%%%%%%%%%%%%%%%%%%%%%%%%%%%%%%%%%%%%%
\author{Arash Azizi}
    \affiliation{Texas A\&M University, College Station, TX 77843}
%%%%%%%%%%%%%%%%%%%%%%%%%%%%%%%%%%%%%%%%%%%%%%%%%%%%%%%%%%%%%%%%%%%%%%%%%%%%

\begin{abstract}
The Unruh effect is notoriously difficult to detect, as it is exponentially overwhelmed by Wigner--Weisskopf spontaneous emission. We show that this fundamental obstacle can be overcome by harnessing multi-detector quantum interference. By preparing a system of three entangled Unruh--DeWitt detectors in a specific W-state, the spontaneous emission channels can be forced to destructively interfere and vanish, thereby "saving" the Unruh signal by coherently silencing this dominant noise. Our central result is the derivation of the condition for complete and simultaneous cancellation of all right- and left-traveling emission modes. We find this requires preparing the detectors in a unique entangled state whose real-valued coefficients are fixed by an elegant geometric constraint, given by a ratio of sines of the logarithms of the detector accelerations. This work establishes multi-detector entanglement as a precision tool for noise cancellation in relativistic quantum settings, offering a new pathway toward the definitive observation of the Unruh signal.
\end{abstract}

\maketitle

%%%%%%%%%%%%%%%%%%%%%%%%%%%%%%%%%%%%%%%%%%%%%%%%%%%%%%%%%%%%%%%%%%%%%%%%%%%%
\section{Introduction} \label{sec:intro}
%%%%%%%%%%%%%%%%%%%%%%%%%%%%%%%%%%%%%%%%%%%%%%%%%%%%%%%%%%%%%%%%%%%%%%%%%%%%

The Unruh effect---the prediction that a uniformly accelerated observer perceives the Minkowski vacuum as a thermal state at temperature $T_U=a/2\pi$---is a cornerstone of quantum field theory in curved spacetime \cite{Unruh1976, Davies1975, Fulling1973}. Closely tied to Hawking radiation \cite{Hawking1974, Hawking1975}, it reshapes our understanding of particles, vacua, and observer dependence \cite{Birrell_Davies1982, Wald1994, Mukhanov2007}. However, a direct experimental observation remains elusive. The most important obstacle is that appreciable Unruh temperatures would require enormous accelerations in direct-detection scenarios (for example, $T_U=1\,\mathrm{K}$ already corresponds to $a\simeq 2.5\times 10^{20}\,\mathrm{m/s^2}$). Even setting that challenge aside (for instance, in effective-acceleration or analogue settings), a second and more immediate obstruction persists: a fundamental signal-to-noise problem.

For a two-level Unruh--DeWitt (UDW) \cite{Unruh1976, Einstein100, Colosi2009Rovelli}  detector ($\ket{g}$ as a ground state and $\ket{e}$ as excited one), the Unruh effect manifests as ($\ket{g}\!\to\!\ket{e}$). However, this "Unruh signal" is accompanied by a much stronger spontaneous emission process ($\ket{e}\!\to\!\ket{g}$). In the eternal, pointlike limit, the rates are related by the Boltzmann factor $e^{2\pi\omega/a}$ \cite{Wigner_Weisskopf1930}, meaning the desired  signal is \emph{exponentially overwhelmed} by this Wigner--Weisskopf (WW) noise. This presents the central experimental challenge: the faint Unruh signal is rendered unobservable by the intrinsically brighter noise.

We show that this fundamental obstacle can be overcome by harnessing multi-detector quantum interference, which, by preparing the detectors in a specific entangled state, allows the spontaneous emission channels to destructively interfere and vanish. To achieve this, we model three UDW detectors prepared in a single-excitation W-state: $\ket{\Psi_i} = (\alpha_1\ket{egg} + \alpha_2\ket{geg} + \alpha_3\ket{gge})\ket{0_M}$. Spontaneous emission from any detector leads to the final state $\ket{ggg}$. The total probability of this spontaneous emission noise is the continuum integral over all Unruh \cite{Unruh1976,UnruhWald1984} modes $\Omega$: $P_{SE} = \int d\Omega | \sum_{k=1}^3 \alpha_k \mathcal{A}_k(\Omega) |^2$. As we show in Sec.~\ref{sec:formalism}, the first-order amplitude for a single detector $k$, $\mathcal{A}_k(\Omega)$, is proportional to a delta function, $\delta(\chi a_k\Omega - \omega_k)$ (see Eq.~\eqref{eq:delta}), where $\chi$ is the parity label (see Eq.~\ref{eq:parity}). This delta function enforces energy conservation, meaning the integrand is zero almost everywhere. Each detector $k$ can \emph{only} emit into a single on-shell frequency $\Omega_k = \omega_k / (\chi a_k)$. The key to our protocol is imposing the resonance condition $\omega_k/a_k = \Lambda$ for all three detectors. This condition forces all detectors, despite their different accelerations, to emit into the exact same on-shell mode, $\Omega = \Lambda/\chi$. This is not a single-mode \emph{approximation}; rather, the physics of the eternal limit \emph{selects} this single, common mode from the continuum. At this one active frequency, the total amplitude factorizes into a common spectral envelope $C(\Lambda)$ and a pure phase sum:
$$
\text{Amp}(\ket{ggg}, \Omega=\Lambda/\chi) \propto C(\Lambda) \times \left[ \sum_{k=1}^3 \alpha_k a_k^{i\chi\Lambda} \right]
$$
This factorization allows us to cancel the total integrated probability $P_{SE}$ by simply setting the "coherent phase sum" in the brackets to zero. To cancel emission into \emph{both} right- and left-traveling modes (for $\chi = \pm 1$, governed by $a_k^{i\Lambda}$ and $a_k^{-i\Lambda}$), we solve a simple linear system. This solution uniquely fixes the state coefficients $\alpha_k$ to be real values given by a compact "sine rule"---our central finding (Eq.~\ref{eq:sine_rule}). Crucially, the $\ket{g}\!\to\!\ket{e}$ absorption signal (the surviving Unruh signal) scales with the conjugate phase and \emph{survives} this cancellation, yielding a clean operational signature.

The goal of this paper is therefore not to verify the thermal detailed-balance criterion, but to solve the prerequisite problem: to achieve the first-ever unambiguous detection of the $\ket{g}\!\to\!\ket{e}$ signal itself. While our protocol, by design, prevents a direct check of detailed balance, it eliminates the dominant spontaneous emission noise, allowing the faint absorption signal to be isolated and observed for the first time.

This multipartite cancellation mechanism is conceptually distinct from \emph{Worldline-Induced Transparency} (WIT), which we derive in a companion article \cite{Azizi2025WIT}. WIT operates through a \emph{single-detector, two-path} interference effect and demonstrates the cancellation of the \emph{Unruh signal} via a similar geometric condition as the present paper. In contrast, the present framework exploits \emph{detector-detector entanglement} among distinct worldlines to cancel the \emph{spontaneous emission noise}. This full, rate-level cancellation of both right- and left-traveling continua requires multipartite resources ($N\!\ge\!3$), leading naturally to the sine-rule constraints derived here.

Our contribution is therefore \emph{theoretical}: in the eternal, pointlike limit we derive exact Unruh-mode transition rules and a closed "sine-rule" constraint that enforces complete RTW/LTW cancellation at the first-order-probability level. Higher-order effects (two-photon channels) are analyzed in \cite{Azizi2025TPE, Azizi2025Dyson-Unruh}; here we focus on first-order rate cancellation, which scales as $g^2$. The residual, un-cancelled noise from second-order amplitudes scales as $g^4$, confirming the robustness of the cancellation in the perturbative regime. These developments motivate translation to analog quantum simulators, where the core result stands independent of platform. A rigorous analysis of finite-time interactions and spatial smearing, while crucial for a complete experimental blueprint, is a separate investigation and is deferred to a future work.

The difficulty of direct detection has motivated analogue-gravity platforms, notably Bose--Einstein condensates (BECs) \cite{Garay2000Sonic, Carusotto2008Numerical_BEC, Lahav2010Sonic_BEC, Jaskula2012Casimir_BEC, Eckel2018Expanding_Universe, Retzker2008Unruh_BEC, Gooding:2020scc, langen2013local, Barcelo2011LRR, Barcelo:2005fc, torres2017rotational}. In relativistic quantum information, complementary interference-based approaches---most prominently Berry-phase probes of acceleration \cite{Martín-Martinez2011Berry}---exploit geometric phases. The present work is complementary to these directions. Rather than accumulating a geometric phase or mapping to an analogue medium, we harness \emph{detector-detector entanglement} to achieve \emph{rate-level} cancellation of WW spontaneous emission.

The remainder of the paper is organized as follows. Section~\ref{sec:formalism} introduces the model, the Unruh-mode expansion, and the mode-resolved transition rules. Section~\ref{sec:emission} assembles the multipartite interference framework: \ref{sec:dualW} treats partial suppression with a dual \(W\) state, \ref{sec:Wstate} presents complete cancellation with a single-excitation \(W\) state, and \ref{sec:RTWLTW} derives simultaneous right/left cancellation and the sine rule. We conclude in Section~\ref{sec:conclusion}. Technical materials are provided in the Appendices: Appendix~\ref{app:Surviving_Unruh} works out the surviving Unruh amplitude in the dual-\(W\) scheme; Appendix~\ref{app:robustness} quantifies sensitivity to acceleration and state-preparation errors; Appendix~\ref{app:3+1} maps the \((1\!+\!1)\)D analysis to the \((3\!+\!1)\)D \(s\)-wave channel; and Appendix~\ref{app:causality} clarifies the RTW/LTW cancellation and its consistency with causality.

%%%%%%%%%%%%%%%%%%%%%%%%%%%%%%%%%%%%%%%%%%%%%%%%%%%%%%%%%%%%%%%%%%%%%%%%%%%%
\section{Formalism}
\label{sec:formalism}
%%%%%%%%%%%%%%%%%%%%%%%%%%%%%%%%%%%%%%%%%%%%%%%%%%%%%%%%%%%%%%%%%%%%%%%%%%%%

We consider three pointlike UDW detectors coupled to a real
scalar field in $(1+1)$ Minkowski spacetime. We work in the \emph{eternal}
and \emph{zero-size} limits (the standard UDW idealization). Finite-time
switching is addressed elsewhere; its main effect is to multiply the on-shell
amplitudes by smooth common envelopes and does not alter the phase-sum
interference mechanism.

\begin{figure}[ht]
    \centering
    \includegraphics[width=1\linewidth]{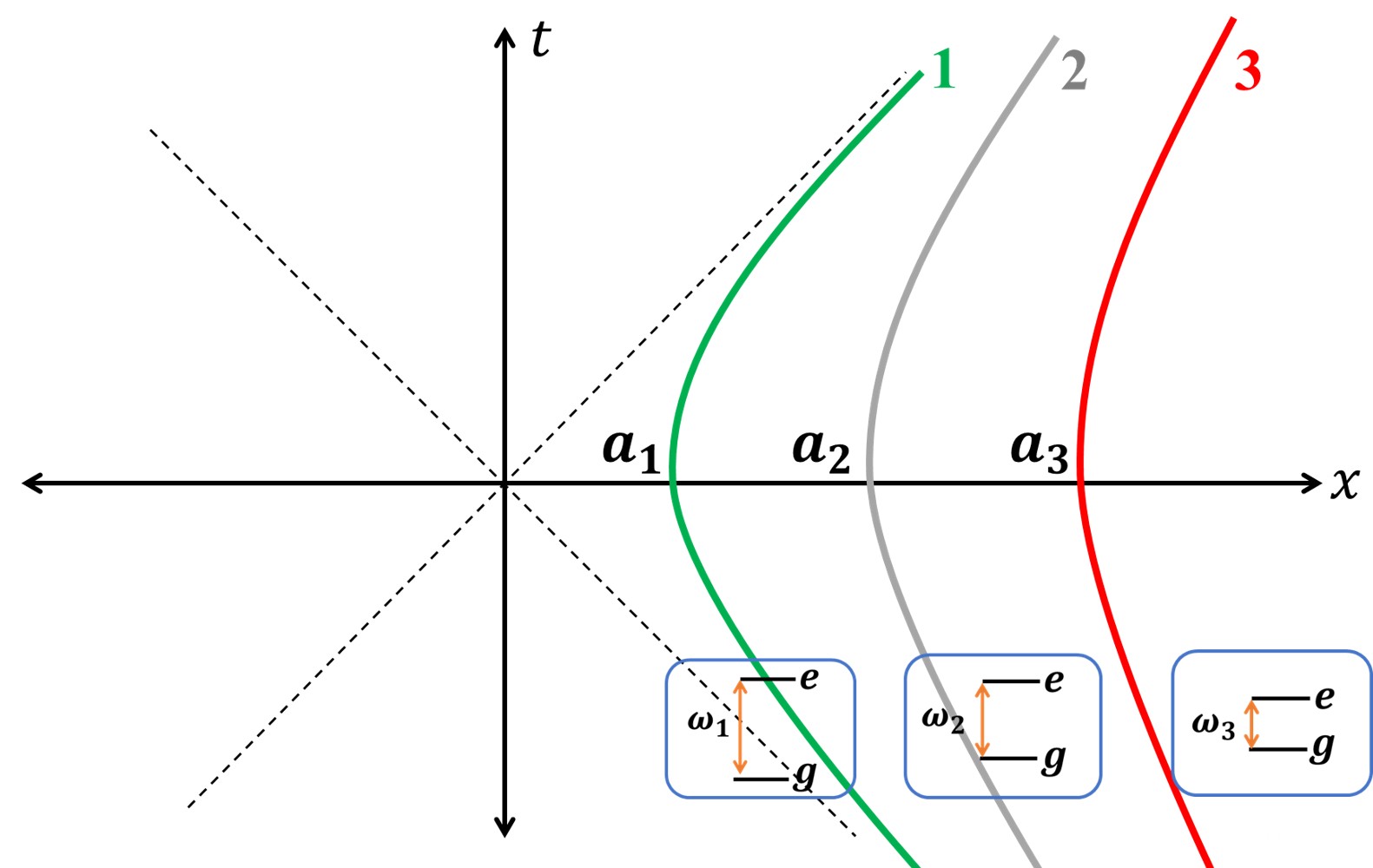}
    \caption{Spacetime diagram of the proposed scenario. Three UDW detectors (1, 2, 3) follow hyperbolic trajectories with distinct proper accelerations $a_1, a_2, a_3$. Each detector is a two-level system with a unique energy gap $\omega_k$. By preparing the detectors in an entangled state and tuning the parameters to satisfy the resonance condition $\frac{\omega_k}{a_k} = \Lambda$, specific field-mediated transitions can be coherently cancelled.}
    \label{fig:setup}
\end{figure}

We use light-cone coordinates
\begin{align}
u \equiv t-x, \qquad v \equiv t+x,
\end{align}
and a parity label \cite{Azizi2025TPE} $\chi=\pm1$ with
\begin{align}
w_\chi \equiv t-\chi x =
\begin{cases}
u, & \chi=+1 \ \text{(right-traveling, RTW)},\label{eq:parity}\\
v, & \chi=-1 \ \text{(left-traveling, LTW)}. 
\end{cases}
\end{align}
Detector $k$ follows a uniformly accelerated trajectory in the right wedge,
\begin{align}
u_k(\tau_k) = -\frac{1}{a_k}e^{-a_k \tau_k}, \qquad
v_k(\tau_k) = \ \frac{1}{a_k}e^{+a_k \tau_k},
\end{align}
with proper time $\tau_k$ and proper acceleration $a_k>0$. Its monopole
operator is
\begin{align}
m_k(\tau_k) \;=\; \sigma_k e^{-i\omega_k \tau_k}
\;+\; \sigma_k^\dagger e^{+i\omega_k \tau_k},
\end{align}
with gap $\omega_k>0$. We use a derivative coupling
\cite{Svidzinsky2021PRL,Svidzinsky2021PRR}, which along the worldline equals
the proper-time directional derivative $u^\mu\partial_\mu\Phi=d\Phi/d\tau_k$:
\begin{align}
H_{I,k}(\tau_k) \;=\;& g \,
\frac{d\Phi\!\big(x_k(\tau_k)\big)}{d\tau_k}\, m_k(\tau_k),\nn\\
%%%%%%%%
U^{(1)} \;=\;& -i \sum_{k=1}^{3}
\int_{-\infty}^{+\infty}\! d\tau_k \; H_{I,k}(\tau_k).
\end{align}
%%%%%%%%%%%%%%%%

We expand the field in Unruh modes \cite{Unruh1976,UnruhWald1984,Azizi2025TPE}
(using the principal branch for $w^{\pm i\Omega}$ and the Heaviside function
$\theta$):
\begin{align}
&\Phi_{\chi}(w) = \int_{-\infty}^{+\infty}\! d\Omega \, \label{Unruh.mode}\\
& \times
\Bigg[ \big(\theta(w)\,f(\Omega)\,w^{+i\Omega}
      +\theta(-w)\,f(-\Omega)\,(-w)^{+i\Omega}\big)\,\mathsf{C}_{\chi,\Omega}
\nn\\
&\quad +\big(\theta(w)\,f(\Omega)\,w^{-i\Omega}
      +\theta(-w)\,f(-\Omega)\,(-w)^{-i\Omega}\big)\,
      \mathsf{C}^\dagger_{\chi,\Omega}\Bigg],
\nn
\end{align}
with
\begin{align}
f(\Omega)=&\frac{e^{-\pi \Omega/2}}{\sqrt{8\pi\,\Omega\,\sinh(\pi\Omega)}},\nn\\
%%%%%%%%
\mathsf{C}^\dagger_{+,\Omega}=&A^\dagger_{\Omega},\quad \qquad
\mathsf{C}^\dagger_{-,\Omega}=B^\dagger_{\Omega},
\end{align}
where $A_{\Omega}$ and $B_{\Omega}$ are annihilation operators for RTW and LTW respectively. The Unruh creators/annihilators obey the bosonic relations
\begin{align}
[A_{\Omega},A_{\Omega'}^\dagger]&=\delta(\Omega-\Omega'),\qquad
[B_{\Omega},B_{\Omega'}^\dagger]=\delta(\Omega-\Omega'),
\end{align}
and act on the Minkowski vacuum as
\begin{align}
A_{\Omega}\ket{0_M}=B_{\Omega}\ket{0_M}=0.
\end{align}
Unruh modes are not restricted in any Rindler wedges; the RTW/LTW labels refer to propagation direction, not confinement to a single Rindler wedge.

Restricting the field to the $k$th UDW's worldline yields
\begin{align}
\Phi_{k,\chi}(\tau_k)
=& \int_{-\infty}^{+\infty}\! d\Omega \;
f(-\chi\Omega)\, a_k^{\,i\Omega} \, e^{\,i \chi a_k \Omega \tau_k} \,
\mathsf{C}^\dagger_{\chi,\Omega} + \text{h.c.},
\end{align}

To first order, the de-excitation (WW / spontaneous emission) and excitation
(Unruh) channels for detector $k$ are
\begin{align}
\ket{e,0_M}
&\ \to\ \sum_{\chi=\pm1}\int_{-\infty}^{+\infty}\! d\Omega\;
\mathcal{A}^{(-)}_{\chi}(\Omega)\,\ket{g;1_{\chi,\Omega}},\nn\\
\ket{g,0_M}
&\ \to\ \sum_{\chi=\pm1}\int_{-\infty}^{+\infty}\! d\Omega\;
\mathcal{A}^{(+)}_{\chi}(\Omega)\,\ket{e;1_{\chi,\Omega}}.
\end{align}
Here the arrow denotes a first-order transition, and the coefficients are the
corresponding matrix elements,
\begin{align}
\mathcal{A}^{(-)}_{\chi}(\Omega)
&\equiv \bra{g;1_{\chi,\Omega}}\,U^{(1)}\,\ket{e;0_M},\nn\\
\mathcal{A}^{(+)}_{\chi}(\Omega)
&\equiv \bra{e;1_{\chi,\Omega}}\,U^{(1)}\,\ket{g;0_M}.
\end{align}
One may use a compact notation, \emph{single sign} $s=\pm1$ distinguishes the process: $s=-1$ (WW/de-excitation) and $s=+1$ (Unruh/excitation). Then,
\begin{align}
\mathcal{A}^{(s)}_{\chi}(\Omega)
=&-i\int_{-\infty}^{\infty}d\tau\;
\frac{d}{d\tau}\Phi(\tau)\,e^{i s\omega\tau} \nn\\
=& -i g \int_{-\infty}^{\infty}d\tau\;
\frac{d}{d\tau}\!\Big[f(-\chi\Omega)\,a^{i\Omega}e^{\,i\chi a\Omega \tau}\Big]\,
e^{i s\omega\tau} \nn\\
=& g\,(\chi a\Omega)\,f(-\chi\Omega)\,a^{i\Omega}\,
\int_{-\infty}^{\infty}d\tau\;e^{\,i(\chi a\Omega+s\omega)\tau}. \nn
\end{align}
Using \(\int_{-\infty}^{\infty}d\tau\,e^{i x\tau}=2\pi\,\delta(x)\) gives
\begin{align}
\mathcal{A}^{(s)}_{\chi}(\Omega)
=&2\pi g\,(\chi a\Omega)\,f(-\chi\Omega)\,a^{i\Omega}\;
\delta(\chi a\Omega+s\omega).
\end{align}
Integrating over \(\Omega\) (i.e., enforcing the
\(\delta\)-constraint) yields the on-shell channel amplitude
\begin{align}
\int_{-\infty}^{\infty}&d\Omega\;\mathcal{A}^{(s)}_{\chi}(\Omega) \ket{1_{\chi,\Omega}} \nn\\
=&\int_{-\infty}^{\infty} d\Omega \,2\pi g\,(\chi a\Omega)\,f(-\chi\Omega)\,a^{i\Omega}\;
\delta(\chi a\Omega+s\omega) \ket{1_{\chi,\Omega}} \nn\\
=&\int_{-\infty}^{\infty} d\Omega \,2\pi g\,(\chi \Omega)\,f(-\chi\Omega)\,a^{i\Omega}\;
\delta\!\Big(\Omega+ s\chi \frac{\omega}a\Big) \ket{1_{\chi,\Omega}} \nn\\
=&-2\pi g\, s\Lambda\,f(s\Lambda)\,a^{-i s\chi\,\Lambda}\,
\ket{1_{\chi,\Omega=-s\chi\Lambda}}, \label{eq:delta}
\end{align}
where we used $\delta(ax)=\frac1{|a|}\delta(x)$ and \(\Lambda\equiv \omega/a\).
Because \(f(s\Lambda)=e^{-s\pi\Lambda/2}\big/\sqrt{8\pi\,\Lambda\,\sinh(\pi\Lambda)}\), this gives explicitly
\begin{align}
\mathcal{A}^{(s)}_{+}
=&-2\pi g\,s\Lambda\,
\frac{e^{-s\frac{\pi\Lambda}{2}}}{\sqrt{8\pi\,\Lambda\,\sinh(\pi\Lambda)}}\;
a^{-i s\Lambda},\nn\\
\mathcal{A}^{(s)}_{-}
=&-2\pi g\,s\Lambda\,
\frac{e^{-s\frac{\pi\Lambda}{2}}}{\sqrt{8\pi\,\Lambda\,\sinh(\pi\Lambda)}}\;
a^{+i s \Lambda}
=\Big(\mathcal{A}^{(s)}_{+}\Big)^{\!*}.
\end{align}
With this notation, excitation vs.\ de-excitation is carried entirely by the
sign $s=\pm1$, and for fixed $s$ the RTW/LTW channels are complex conjugates.
It is convenient to package the spectral factors as
\begin{align}
I(\omega,a) \equiv \mathcal{A}^{(-)}_{+}
= 2\pi g\,\frac{\omega}a\;
   \frac{e^{+\frac{\pi\omega}{2a}}}{\sqrt{8\pi\,\frac{\omega}a\,\sinh(\pi\frac{\omega}a)}}\;
   a^{\,i\frac{\omega}a}.
\label{eq:I-spectral}
\end{align}
The single-detector transition rules then read
\begin{align}
&\ket{e_k,0_M}
\ \to\  \label{eq:rule_emission}\\[-2pt]
&\hspace{1em}
\Big(\, I(\omega_k,a_k)\,A^\dagger_{+\Lambda_k}
      \;+\; I^{*}(\omega_k,a_k)\,B^\dagger_{-\Lambda_k}\,\Big)
\ket{g_k,0_M}, \nn\\[4pt]
&\ket{g_k,0_M}
\ \to\ \label{eq:rule_absorption}\\[-2pt]
&\hspace{1em}
\Big(\, I(-\omega_k,a_k)\,A^\dagger_{-\Lambda_k}
      \;+\; I^{*}(-\omega_k,a_k)\,B^\dagger_{+\Lambda_k}\,\Big)
\ket{e_k,0_M}, \nn
\end{align}
with $\Lambda_k\equiv\omega_k/a_k$. From \eqref{eq:I-spectral} one immediately
obtains detailed balance,
\begin{align}
\frac{|I(\omega,a)|^2}{|I(-\omega,a)|^2}=e^{\frac{2\pi\omega}{a}}.
\end{align}

%%%%%%%%%%%%%%%%%%%%%%%%%%%%%%%%%%%%%%%%%%%%%%%%%%%%%%%%%%%%%%%%%%%%%%%%%%%%
\section{Emission Suppression with a Dual W-State} \label{sec:emission}
%%%%%%%%%%%%%%%%%%%%%%%%%%%%%%%%%%%%%%%%%%%%%%%%%%%%%%%%%%%%%%%%%%%%%%%%%%%%

%%%%%%%%%%%%%%%%%%%%%%%%%%%%%%%%%%%%%%%%%%%%%%%%%%%%%%%%%%%%%%%%%%%%%%%%%%%%
\subsection{Partial Emission Suppression with a Dual W-State} \label{sec:dualW}
%%%%%%%%%%%%%%%%%%%%%%%%%%%%%%%%%%%%%%%%%%%%%%%%%%%%%%%%%%%%%%%%%%%%%%%%%%%%

Our first strategy aims to suppress the dominant spontaneous emission channels. We consider three detectors prepared in a ``dual'' W-state with two excitations:
\begin{align}
\ket{\Psi_i} = \left( \alpha_1 \ket{gee} + \alpha_2 \ket{ege} + \alpha_3 \ket{eeg} \right)\ket{0_M}.
\label{eq:initial_state_dual}
\end{align}

The collective Unruh signal, exciting the single ground-state detector in each term, has an amplitude:
\begin{align}
\text{Amp}(\ket{eee}) = \sum_{k=1}^{3} \alpha_k\, I(-\omega_k, a_k).
\label{eq:amp_unruh}
\end{align}
Under the resonance condition \(\omega_k/a_k\equiv\Lambda\) we write
\begin{align}
I(\omega_k,a_k)=C_{+}(\Lambda)\,a_k^{i\Lambda},\qquad
I(-\omega_k,a_k)=C_{-}(\Lambda)\,a_k^{-i\Lambda},
\end{align}
with \(C_{\pm}(\Lambda)\) independent of \(k\).

Spontaneous emission leads to three possible final atomic states, with amplitudes
\begin{align}
\text{Amp}(\ket{gge}) =&\ \alpha_1 I(\omega_2, a_2) + \alpha_2 I(\omega_1, a_1), \nn\\
\text{Amp}(\ket{geg}) =&\ \alpha_1 I(\omega_3, a_3) + \alpha_3 I(\omega_1, a_1),  \nn\\
\text{Amp}(\ket{egg}) =&\ \alpha_2 I(\omega_3, a_3) + \alpha_3 I(\omega_2, a_2). \label{eq:amp_ww3}
\end{align}
Imposing \(\text{Amp}(\ket{gge})=\text{Amp}(\ket{geg})=0\) gives
\begin{align}
\frac{\alpha_2}{\alpha_1} = -\Big(\frac{a_2}{a_1}\Big)^{i\Lambda},\qquad
\frac{\alpha_3}{\alpha_1} = -\Big(\frac{a_3}{a_1}\Big)^{i\Lambda}.
\end{align}
Substituting into \(\text{Amp}(\ket{egg})=0\) yields
\[
\text{Amp}(\ket{egg})=-2\,C_{+}(\Lambda)\,\alpha_1\,
\Big(\frac{a_2 a_3}{a_1}\Big)^{i\Lambda}=0
\ \Rightarrow\ \alpha_1=0,
\]
hence \(\alpha_2=\alpha_3=0\). Therefore, it is impossible to simultaneously cancel all three spontaneous emission channels with this state (this already holds for a single parity; cancellation must in any case be enforced separately for RTW and LTW).

However, by enforcing only the first two conditions to nullify the amplitudes for \(\ket{gge}\) and \(\ket{geg}\), the Unruh signal survives. Using \(I(-\omega_k,a_k)=C_{-}(\Lambda)\,a_k^{-i\Lambda}\),
\begin{align}
\text{Amp}(\text{Unruh})
=& C_{-}(\Lambda)\!\sum_{k=1}^{3}\alpha_k a_k^{-i\Lambda} \\
%%%%%%%%%
=& -\,C_{-}(\Lambda)\,\alpha_1\,a_1^{-i\Lambda}
\ \propto\ -\,\alpha_1\,a_1^{-i\Lambda}\neq 0. \nn
\end{align}
Thus, by nullifying two of the three dominant noise channels, the collective Unruh signal remains robustly detectable.

%%%%%%%%%%%%%%%%%%%%%%%%%%%%%%%%%%%%%%%%%%%%%%%%%%%%%%%%%%%%%%%%%%%%%%%%%%%%
\subsection{Complete (single-parity) Emission Suppression with a W-State} \label{sec:Wstate}
%%%%%%%%%%%%%%%%%%%%%%%%%%%%%%%%%%%%%%%%%%%%%%%%%%%%%%%%%%%%%%%%%%%%%%%%%%%%

An alternative strategy uses a standard W-state with a single excitation:
\begin{align}
\ket{\Psi_i} = \left( \alpha_1 \ket{egg} + \alpha_2 \ket{geg} + \alpha_3 \ket{gge} \right) \ket{0_M}.
\label{eq:initial_state_w}
\end{align}

Here, spontaneous emission from any branch populates the state $\ket{ggg}$. This is the undesired spontaneous emission channel to be cancelled. Its (single-parity) amplitude is
\begin{align}
\text{Amp}(\ket{ggg}) = \alpha_1 I(\omega_1, a_1) + \alpha_2 I(\omega_2, a_2) + \alpha_3 I(\omega_3, a_3).
\label{eq:amp_ggg}
\end{align}
Under the resonance condition $\omega_k/a_k=\Lambda$, write
\begin{align}
I(\omega_k,a_k)=C_{+}(\Lambda)\,a_k^{i\Lambda},\qquad
I(-\omega_k,a_k)=C_{-}(\Lambda)\,a_k^{-i\Lambda},
\end{align}
with $C_{\pm}(\Lambda)$ independent of $k$. The cancellation condition for the RTW (right-traveling) channel is:
\begin{align}
\alpha_1 a_1^{i\Lambda} + \alpha_2 a_2^{i\Lambda} + \alpha_3 a_3^{i\Lambda} = 0.
\label{eq:w_cancel_condition}
\end{align}
This single constraint is not overly restrictive and permits non-trivial solutions for the coefficients $\alpha_k$.

Under \eqref{eq:w_cancel_condition}, the Unruh signals—which lead to two-excitation final states—\emph{generically} survive because absorption scales as $a_k^{-i\Lambda}$ (different phase law). They would vanish only if the \emph{conjugate} condition $\sum_k \alpha_k a_k^{-i\Lambda}=0$ were also (accidentally) satisfied; the simultaneous case is analyzed next.

For example, the amplitude for the transition to $\ket{ege}$ is
\begin{align}
\text{Amp}(\ket{ege}) =& \alpha_1 I(-\omega_3, a_3) + \alpha_3 I(-\omega_1, a_1) \nn\\
%%%%%%%%%%
=& C_{-}(\Lambda)\,\big(\alpha_1 a_3^{-i\Lambda} + \alpha_3 a_1^{-i\Lambda}\big).
\end{align}
Using \eqref{eq:w_cancel_condition},
\[
\alpha_1 a_1^{i\Lambda} + \alpha_3 a_3^{i\Lambda} = -\alpha_2 a_2^{i\Lambda}\ ,
\]
then 
\begin{align}
    \alpha_1 a_3^{-i\Lambda} + \alpha_3 a_1^{-i\Lambda}
= -\,\alpha_2\,a_1^{-i\Lambda} a_2^{i\Lambda} a_3^{-i\Lambda}
\end{align}
so
\begin{align}
\text{Amp}(\ket{ege}) =& -\,C_{-}(\Lambda)\,\alpha_2\,a_1^{-i\Lambda} a_2^{i\Lambda} a_3^{-i\Lambda} \nn\\
%%%%%
     \propto & -\,\alpha_2\,a_1^{-i\Lambda} a_2^{i\Lambda} a_3^{-i\Lambda}\neq 0.
\end{align}
Thus one can completely remove the RTW spontaneous emission  while preserving the Unruh signal (unless the LTW cancellation is also imposed).

%%%%%%%%%%%%%%%%%%%%%%%%%%%%%%%%%%%%%%%%%%%%%%%%%%%%%%%%%%%%%%%%%%%%%%%%%%%%
\subsection{Simultaneous RTW and LTW Cancellation} \label{sec:RTWLTW}
%%%%%%%%%%%%%%%%%%%%%%%%%%%%%%%%%%%%%%%%%%%%%%%%%%%%%%%%%%%%%%%%%%%%%%%%%%%%

We now investigate the most stringent conditions required to cancel spontaneous emission into both RTW and LTW modes. For fixed process $s$, the RTW and LTW amplitudes are complex conjugates; cancelling both requires satisfying a condition and its complex conjugate.

\begin{table}[t]
    \caption{Summary of spontaneous-emission cancellation conditions.}
    \label{tab:cancellation_schemes_elegant}
    \renewcommand{\arraystretch}{1.5}
    \setlength{\tabcolsep}{6pt}
    \centering
    \begin{tabular}{@{}ll@{}}
    \toprule
    \textbf{Cancellation Target} & \textbf{Condition} \\
    \midrule
    \multicolumn{2}{l}{\textit{Initial State: Dual W-State}} \\
    \addlinespace
    Partial RTW  & $\displaystyle \frac{\alpha_2}{\alpha_1} = -\left(\frac{a_2}{a_1}\right)^{i\Lambda}$ \\
    \addlinespace
    Partial RTW \& LTW  & $\displaystyle \Lambda \ln\!\left(\frac{a_k}{a_j}\right) = m\pi,\ \ m\in\mathbb{Z}$ \\
    \midrule
    \multicolumn{2}{l}{\textit{Initial State: Standard W-State}} \\
    \addlinespace
    Complete RTW  & $\displaystyle \sum_{k=1}^3 \alpha_k a_k^{i\Lambda} = 0$ \\
    \addlinespace
    Complete RTW \& LTW  & $\displaystyle -\frac{\alpha_i}{\alpha_j} =
      \frac{\sin\!\big(\Lambda \ln\frac{a_j}{a_k}\big)}{\sin\!\big(\Lambda \ln\frac{a_i}{a_k}\big)}$ \\
    \bottomrule
    \end{tabular}
\end{table}

%%%%%%%%%%%%%%%%%%%%%%%%%%%%%%%%%%%%
\begin{figure*}[t]
    \centering
    \includegraphics[width=\textwidth]{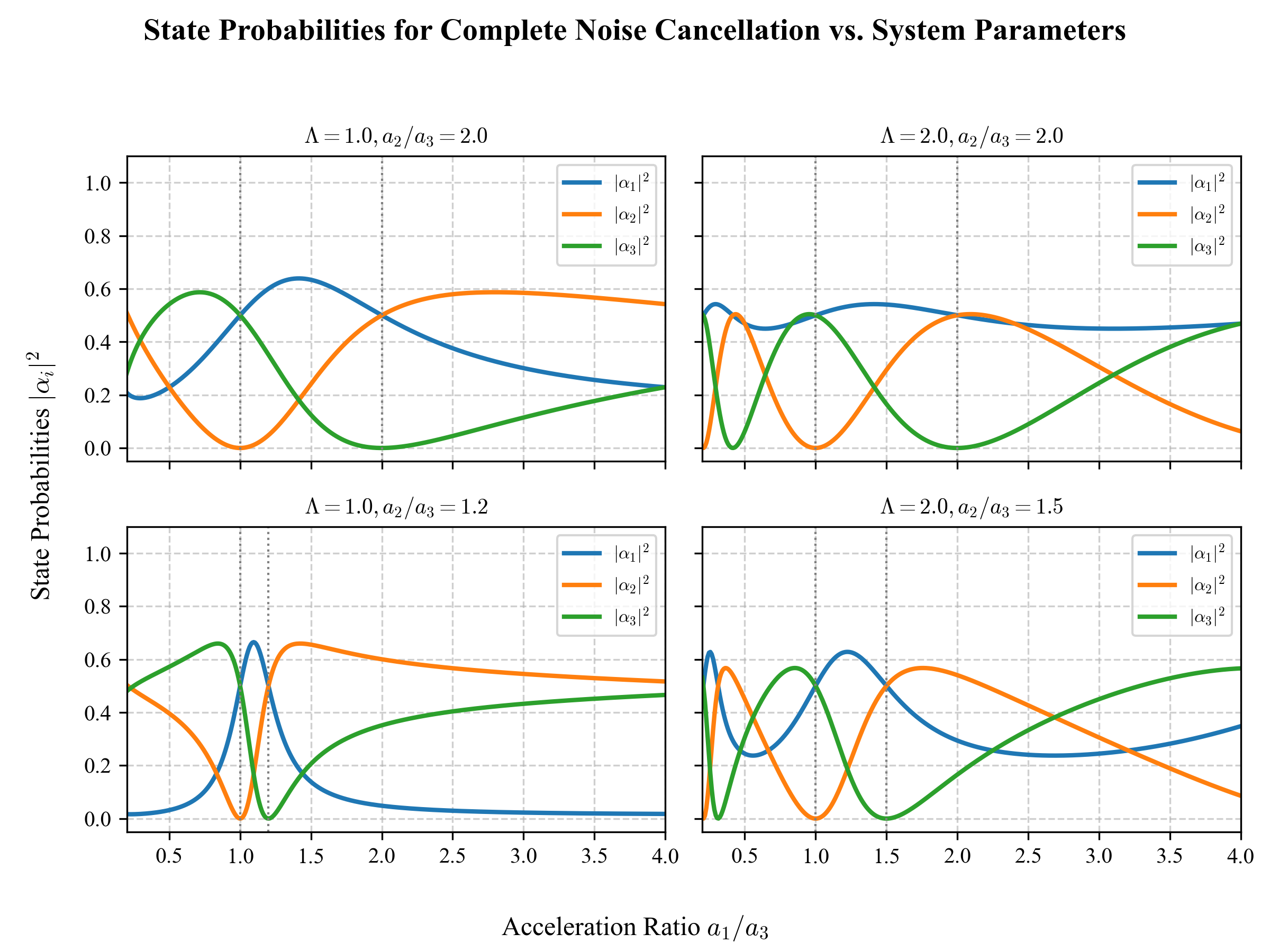}
    \caption{The required probabilities $|\alpha_i|^2$ for the entangled W-state that achieves complete spontaneous emission cancellation, plotted as a function of the acceleration ratio $a_1/a_3$. Each subplot shows the solution for a different set of fixed physical parameters $(\Lambda, a_2/a_3)$, demonstrating how the required quantum state depends on both the resonance condition and the geometric configuration of the detectors. The plots show that increasing $\Lambda$ leads to more rapid oscillations, indicating a higher sensitivity to the system's geometry.}
    \label{fig:probabilities_grid}
\end{figure*}
%%%%%%%%%%%%%%%%%%%%%%%%%%%%%%%%%%%%

For the dual W-state, cancelling a given emission channel (e.g., $\ket{gge}$) in both modes requires
$\left(\frac{a_2}{a_1}\right)^{i\Lambda}=\left(\frac{a_2}{a_1}\right)^{-i\Lambda}$, i.e.
\begin{align}
\Lambda \ln\!\left(\frac{a_2}{a_1}\right) = m\pi,\quad m\in\mathbb{Z}.
\end{align}

For the standard W-state, to cancel $\ket{ggg}$ in both modes we require
\begin{align}
\sum_{k=1}^3 \alpha_k a_k^{i\Lambda} = 0, \qquad \quad
\sum_{k=1}^3 \alpha_k a_k^{-i\Lambda} = 0.
\end{align}
Writing $\theta_k\equiv \Lambda\ln a_k$, these are equivalent to
$\sum_k \alpha_k \cos\theta_k=0$ and $\sum_k \alpha_k \sin\theta_k=0$.
For the \emph{generic non-degenerate} case (no $\theta$ differences equal to $m\pi$), the solution space is one-dimensional (unique up to a global phase). Choosing $\alpha_k$ real (up to a common phase) gives the symmetric ratios
\begin{align}
 -\frac{\alpha_i}{\alpha_j}
 = \frac{\sin\!\big(\theta_j-\theta_k\big)}{\sin\!\big(\theta_i-\theta_k\big)}
 = \frac{\sin\!\Big(\Lambda \ln\frac{a_j}{a_k}\Big)}{\sin\!\Big(\Lambda \ln\frac{a_i}{a_k}\Big)},
\end{align}
for any permutation $\{i,j,k\}=\{1,2,3\}$. With normalization, the probabilities are
\begin{align}
|\alpha_i|^2=
    \frac{
        \sin^2 \!\Big( \Lambda \ln \frac{a_j}{a_k} \Big)
    }{
        \sin^2 \!\Big( \Lambda \ln \frac{a_1}{a_2} \Big)
        +\sin^2 \!\Big( \Lambda \ln \frac{a_2}{a_3} \Big)
        +\sin^2 \!\Big( \Lambda \ln \frac{a_3}{a_1} \Big)
    }, \label{eq:sine_rule}
\end{align}
which sum to unity by construction. Degenerate cases (e.g., $\Lambda\ln(a_p/a_q)=m\pi$) either
make the solution non-unique or preclude simultaneous RTW/LTW cancellation for generic $\alpha$.

%%%%%%%%%%%%%%%%%%%%%%%%%%%%%%%%%%%%%%%%%%%%%%%%%%%%%%%%%%%%%%%%%%%%%%%%%%%%
\section{Conclusion}   \label{sec:conclusion}
%%%%%%%%%%%%%%%%%%%%%%%%%%%%%%%%%%%%%%%%%%%%%%%%%%%%%%%%%%%%%%%%%%%%%%%%%%%%

The direct observation of any Unruh-related physics is fundamentally challenged by an exponentially dominant spontaneous emission noise. In this work, we have designed a quantum interference protocol using a three-detector entangled $W$-state to coherently cancel this dominant noise. Our central result is the derivation of the ``sine rule'' (Eq.~\ref{eq:sine_rule}): a specific, geometric constraint on the entangled state coefficients that simultaneously nullifies the spontaneous emission amplitudes for both right- and left-traveling modes. This cancellation isolates the faint, acceleration-induced absorption ($\ket{g} \to \ket{e}$) transition, transforming it from an obscured process into a clean, operational signal.

This protocol, by design, achieves this clean signal by silencing the de-excitation channel, thus its goal is the unambiguous \emph{detection} of the Unruh signal itself, not a verification of the thermal detailed-balance criterion. We have shown that this mechanism is robust. While presented in (1+1)D, our analysis targets the monopole (s-wave) interaction, making the principle directly applicable to the relevant physics in (3+1)D (App.~\ref{app:3+1}). Furthermore, the cancellation is resilient to the small state-preparation and acceleration-control errors inherent in any realistic experiment (App.~\ref{app:robustness}).

This work establishes multi-detector entanglement as a precision tool for noise cancellation in relativistic quantum settings. By demonstrating a method to completely silence the dominant spontaneous emission, we offer a new pathway toward the first definitive observation of the Unruh signal.

\appendix

%%%%%%%%%%%%%%%%%%%%%%%%%%%%%%%%%%%%%%%%%%%%%%%%%%%%%%%%%%%%%%%%%%
\section{Derivation of the Surviving Unruh Signal Amplitude} \label{app:Surviving_Unruh}
%%%%%%%%%%%%%%%%%%%%%%%%%%%%%%%%%%%%%%%%%%%%%%%%%%%%%%%%%%%%%%%%%%

We give a step-by-step derivation of the surviving Unruh (excitation) amplitude
for the dual-$W$ initial state under the partial emission-suppression scheme.
For clarity we present the calculation for a single propagation sector (e.g.,
RTW); the LTW contribution is the complex conjugate and follows identically.

\subsection{Initial state and amplitudes}
We begin with
\begin{equation}
\ket{\Psi_i}=\Big(\alpha_1\ket{gee}+\alpha_2\ket{ege}+\alpha_3\ket{eeg}\Big)\ket{0_M}.
\end{equation}
The collective Unruh signal corresponds to $\ket{eee}$, with amplitude
\begin{equation}
\text{Amp}(\ket{eee})
=\alpha_1 I(-\omega_1,a_1)+\alpha_2 I(-\omega_2,a_2)+\alpha_3 I(-\omega_3,a_3).
\label{eq:sm_unruh_amp_start}
\end{equation}
The two spontaneous-emission channels we cancel are those populating
$\ket{gge}$ and $\ket{geg}$:
\begin{align}
\text{Amp}(\ket{gge})&=\alpha_1 I(\omega_2,a_2)+\alpha_2 I(\omega_1,a_1),\nn\\
\text{Amp}(\ket{geg})&=\alpha_1 I(\omega_3,a_3)+\alpha_3 I(\omega_1,a_1).
\label{eq:amp_ww2}
\end{align}

\subsection{Imposing the cancellation conditions}
Impose the resonance condition $\omega_k/a_k\equiv\Lambda$. Then
\begin{align}
I(\omega_k,a_k)&=\tilde I(\Lambda)\,a_k^{i\Lambda},\nn\\
I(-\omega_k,a_k)&=\tilde I(-\Lambda)\,a_k^{-i\Lambda},
\end{align}
with $\tilde I(\pm\Lambda)$ independent of $k$. Setting \eqref{eq:amp_ww2} to
zero yields
\begin{align}
\alpha_2&=-\alpha_1\Big(\frac{a_2}{a_1}\Big)^{i\Lambda},\label{eq:sm_alpha2_sol}\\
\alpha_3&=-\alpha_1\Big(\frac{a_3}{a_1}\Big)^{i\Lambda}.\label{eq:sm_alpha3_sol}
\end{align}

\subsection{Calculating the final Unruh amplitude}
Substituting \eqref{eq:sm_alpha2_sol}--\eqref{eq:sm_alpha3_sol} into
\eqref{eq:sm_unruh_amp_start} gives

\begin{align}
\text{Amp}(\ket{eee})
=&\tilde I(-\Lambda)\Big(\alpha_1 a_1^{-i\Lambda}+\alpha_2 a_2^{-i\Lambda}+\alpha_3 a_3^{-i\Lambda}\Big)\nn\\
=&\tilde I(-\Lambda)\Big[\alpha_1 a_1^{-i\Lambda}
-\alpha_1\Big(\frac{a_2}{a_1}\Big)^{i\Lambda}a_2^{-i\Lambda} \nn\\
%%%
&
-\alpha_1\Big(\frac{a_3}{a_1}\Big)^{i\Lambda}a_3^{-i\Lambda}\Big]\nn\\
=&\tilde I(-\Lambda)\Big[\alpha_1 a_1^{-i\Lambda}-\alpha_1 a_1^{-i\Lambda}-\alpha_1 a_1^{-i\Lambda}\Big]\nn\\
=&-\,\tilde I(-\Lambda)\,\alpha_1 a_1^{-i\Lambda}\ \propto\ -\alpha_1 a_1^{-i\Lambda}\neq 0.
\end{align}

%%%%%%%%%%%%%%%%%%%%%%%%%%%%%%%%%%%%%%%%%%%%%%%%%%%%%%%%%%%%%%%%%%
\section{Robustness Analysis} \label{app:robustness}
%%%%%%%%%%%%%%%%%%%%%%%%%%%%%%%%%%%%%%%%%%%%%%%%%%%%%%%%%%%%%%%%%%

Here, we analyze the stability of the spontaneous-emission cancellation against realistic imperfections. The dominant requirements are: (i) preparation of the target entangled $W$-state, and (ii) maintenance of the resonance manifold $\omega_k/a_k=\Lambda$ that makes the emission amplitudes share a common on-shell mode and hence interfere.

A central point is that, in the strict eternal limit, exact resonance is required for different detectors to populate the \emph{same} on-shell Unruh mode. In practice, any experimental realization employs finite-time interactions (or an equivalent finite bandwidth), so that the energy-conservation $\delta$-function is replaced by a narrow spectral envelope. In this setting, small deviations from perfect resonance do not destroy interference discontinuously; rather, they appear as small residual amplitudes controlled by the overlap of the broadened on-shell envelopes. In what follows we quantify the leading sensitivity in a simple and conservative way.

%%%%%%%%%%%%%%%%%%%%%%%%%%%%%%%%%%%%%%%%%%%%%%%%%%%%%%%%%%%%%%%%%%
\subsection{Perturbations in Detector Acceleration} \label{app:robustness-acc}
%%%%%%%%%%%%%%%%%%%%%%%%%%%%%%%%%%%%%%%%%%%%%%%%%%%%%%%%%%%%%%%%%%

We start from the complete cancellation condition for the standard $W$-state in a single propagation sector (e.g., RTW),
\begin{equation}
\text{Amp}(\ket{ggg})_{\text{RTW}} \propto \sum_{k=1}^3 \alpha_k a_k^{i\Lambda}=0,
\label{eq:dark_rtw}
\end{equation}
where $\Lambda$ is the common resonance parameter defined by $\omega_k/a_k=\Lambda$.

In an experimental platform, the most relevant control knob is typically the effective acceleration $a_k$. A perturbation $a_k\to a_k+\delta a_k$ has two distinct effects: it changes the phase factor $a_k^{i\Lambda}$ and, unless compensated, it also shifts the resonance parameter $\Lambda_k=\omega_k/a_k$. To isolate the multipartite interference sensitivity, we first analyze the \emph{on-resonance} case in which $\Lambda$ is held fixed (for instance, by tuning $\omega_k$ together with $a_k$ so that $\omega_k/a_k=\Lambda$ continues to hold). The remaining imperfection is then a phase error in the factors $a_k^{i\Lambda}$.

With $\Lambda$ fixed, the residual RTW amplitude becomes
\begin{equation}
\delta_a \propto \sum_{k=1}^3 \alpha_k\,(a_k+\delta a_k)^{i\Lambda}.
\end{equation}
For small relative perturbations $|\delta a_k|/a_k\ll 1$, we expand
\begin{align}
(a_k+\delta a_k)^{i\Lambda}
&=a_k^{i\Lambda}\Big(1+\frac{\delta a_k}{a_k}\Big)^{i\Lambda}\nn\\
&\approx a_k^{i\Lambda}\Big(1+i\Lambda\frac{\delta a_k}{a_k}\Big),
\end{align}
so that
\begin{align}
\delta_a
&\propto \sum_{k=1}^3 \alpha_k\Big(a_k^{i\Lambda}+i\Lambda\,a_k^{i\Lambda}\frac{\delta a_k}{a_k}\Big)\nn\\
&\propto i\Lambda \sum_{k=1}^3 \alpha_k\,a_k^{i\Lambda}\frac{\delta a_k}{a_k},
\label{eq:sm_residual_amp_fixedLambda}
\end{align}
where we used the ideal cancellation condition \eqref{eq:dark_rtw} to drop the zeroth-order term. The emitted power scales as $|\delta_a|^2$. A simple norm bound follows from Cauchy--Schwarz:
\begin{align}
|\delta_a|
&\lesssim |\Lambda| \sum_{k=1}^3 |\alpha_k|\Big|\frac{\delta a_k}{a_k}\Big|\nn\\
&\le |\Lambda| \sqrt{3}\,\max_k\Big|\frac{\delta a_k}{a_k}\Big|,
\label{eq:bound_residual}
\end{align}
using $\sum_k|\alpha_k|^2=1$. If the coefficients are real up to a common overall phase, the LTW residual amplitude is the complex conjugate of the RTW one for fixed process $s$, hence it has the same magnitude.

If, instead, the gaps $\omega_k$ are held fixed while $a_k$ fluctuates, then $\Lambda_k=\omega_k/a_k$ also fluctuates. In the strict eternal limit this shifts the on-shell frequencies selected by the energy-conservation condition and can reduce interference. In realistic finite-time (finite-bandwidth) settings, this effect is controlled by the overlap of the corresponding spectral envelopes and can be made parametrically small whenever the resonance mismatch $|\delta\Lambda_k|=|\Lambda|\,|\delta a_k|/a_k$ remains well within the envelope width. We leave a platform-specific envelope analysis to future work.

%%%%%%%%%%%%%%%%%%%%%%%%%%%%%%%%%%%%%%%%%%%%%%%%%%%%%%%%%%%%%%%%%%
\subsection{Imperfections in State Preparation} \label{app:robustness-state}
%%%%%%%%%%%%%%%%%%%%%%%%%%%%%%%%%%%%%%%%%%%%%%%%%%%%%%%%%%%%%%%%%%

A second source of error is imperfect preparation of the entangled $W$-state. Modeling the prepared pure state by a fidelity decomposition,
\begin{equation}
\ket{\Psi_{\text{prepared}}}
=\sqrt{F}\,\ket{\Psi_W}+\sqrt{1-F}\,\ket{\Psi_{\text{err}}},
\end{equation}
with $\bra{\Psi_W}\ket{\Psi_{\text{err}}}=0$, the spontaneous-emission amplitude from the ideal component vanishes by construction. Any residual emission arises from the orthogonal component and therefore scales as $\sqrt{1-F}$ in amplitude and as $(1-F)$ in probability (power). Thus, for $F=0.99$, the residual spontaneous-emission probability contributed by this error channel is suppressed by a factor $\sim 10^{-2}$ relative to a comparable non-dark preparation (up to order-unity factors determined by the specific structure of $\ket{\Psi_{\text{err}}}$).

%%%%%%%%%%%%%%%%%%%%%%%%%%%%%%%%%%%%%%%%%%%%%%%%%%%%%%%%%%%%%%%%%%
\section{Generalization to (3+1) Dimensions and Field Mode Structure} \label{app:3+1}
%%%%%%%%%%%%%%%%%%%%%%%%%%%%%%%%%%%%%%%%%%%%%%%%%%%%%%%%%%%%%%%%%%

Here we clarify how the cancellation mechanism interfaces with a more realistic
\((3+1)\)D field. The mode-resolved derivation in the main text was carried out
in a \((1+1)\)D massless scalar model, for which the field factorizes into two
chiral sectors and the on-shell selection in the eternal limit isolates a
single frequency in each sector. In \((3+1)\)D free space, by contrast, each
frequency is accompanied by an additional degeneracy label (e.g., angles or
transverse momentum), so the statement ``cancellation of an on-shell mode''
must be understood \emph{channel by channel} with respect to these extra labels.

A pointlike monopole UDW detector still has a response determined by the field
two-point function evaluated along its worldline, and the usual Unruh
thermality persists in \((3+1)\)D. However, for a genuinely \((3+1)\)D
continuum the detector couples to a family of degenerate modes at fixed
frequency, and the corresponding mode functions along accelerated worldlines
generally carry additional label-dependent factors beyond the pure phase
\(a_k^{\pm i\Lambda}\). As a result, an exact rate-level cancellation of
spontaneous emission in \((3+1)\)D requires enforcing the destructive
interference condition for each relevant mode label (or, equivalently, for
each independent channel that contributes to the response).

The \((1+1)\)D analysis therefore applies \emph{exactly} to settings in which
the detector couples predominantly to a \emph{single effective channel}---for
example, an effectively one-dimensional environment (waveguide/circuit/cavity
or any platform with a single propagating mode), or a deliberately engineered
projection onto a reduced sector. In such single-channel realizations, the
same phase-sum conditions derived in the main text carry over unchanged, and
the sine-rule state preparation produces complete first-order spontaneous-emission
darkness in both chiral sectors of that effective theory. A detailed mapping
for a specific experimental platform is model-dependent and is left for future
work.

%%%%%%%%%%%%%%%%%%%%%%%%%%%%%%%%%%%%%%%%%%%%%%%%%%%%%%%%%%%%%%%%%%%
\section{Physical Interpretation of Simultaneous RTW/LTW Cancellation and Causality} \label{app:causality}
%%%%%%%%%%%%%%%%%%%%%%%%%%%%%%%%%%%%%%%%%%%%%%%%%%%%%%%%%%%%%%%%%%%

We now comment on the physical meaning of simultaneous RTW/LTW cancellation
and its consistency with causality. The key point is that the cancellation is
an \emph{initial-state} interference effect: the detectors are prepared so that
their combined first-order source for the relevant on-shell field excitations
vanishes. No signal needs to propagate between detectors, and no information
needs to be exchanged across causally disconnected regions.

It is also important not to conflate the RTW/LTW propagation labels with
restriction to a particular Rindler wedge. The Unruh-mode operators used in
the main text create global field excitations (analytic across the horizon)
whose decomposition into wedge-localized Rindler excitations reflects the
entanglement structure of the Minkowski vacuum. In that sense, cancelling both
RTW and LTW emission corresponds to decoupling the detector system from the
two correlated chiral sectors that appear naturally in the Unruh-mode
description.

At the level of amplitudes, simultaneous first-order cancellation for
spontaneous emission imposes
\begin{align}
\sum_{k=1}^3 \alpha_k a_k^{i\Lambda}=0,\qquad
\sum_{k=1}^3 \alpha_k a_k^{-i\Lambda}=0,
\end{align}
which are conjugate constraints once the coefficients are chosen real up to a
common overall phase (as is appropriate for a physical $W$-state with no
relative dynamical phases). The resulting ``sine rule'' follows from a simple
geometric interpretation: the complex equation
\(\sum_k \alpha_k e^{i\theta_k}=0\), with \(\theta_k\equiv \Lambda\ln(a_k/a_0)\),
states that the three phasors \(\alpha_k e^{i\theta_k}\) form a closed
triangle in the complex plane. Applying the law of sines to this triangle
immediately yields the ratios of coefficients in terms of
\(\sin(\theta_i-\theta_j)=\sin\!\big(\Lambda\ln(a_i/a_j)\big)\), producing the
unique (up to normalization and a global phase) real-valued solution quoted in
the main text.

%%%%%%%%%%%%%%%%%%%%%%%%%%%%%%%%%%%%%%%%%%%%%%%%%%%%%%%%%%%%%%%%%%%%%%%%%%%%
\bibliographystyle{apsrev4-2} 
\bibliography{UnruhRef}

@article{Wigner_Weisskopf1930,
  author = "V. Weisskopf and E. Wigner",
  title = "{{Berechnung der natürlichen Linienbreite auf Grund der Diracschen Lichttheorie}}",
  journal = "Zeitschrift für Physik",
  volume = "63",
  number = "1",
  pages = "54--73",
  year = "1930",
  doi = "10.1007/BF01336768",
  issn = "0044-3328",
  url = "https://doi.org/10.1007/BF01336768"
}

@article{Unruh1976,
  title = {{Notes on black-hole evaporation}},
  author = {Unruh, W. G.},
  journal = {Phys. Rev. D},
  volume = {14},
  issue = {4},
  pages = {870--892},
  numpages = {0},
  year = {1976},
  month = {Aug},
  publisher = {American Physical Society},
  doi = {10.1103/PhysRevD.14.870},
  url = {https://link.aps.org/doi/10.1103/PhysRevD.14.870}
}

@article{Fulling1973,
  title = {{Nonuniqueness of Canonical Field Quantization in Riemannian Space-Time}},
  author = {Fulling, Stephen A.},
  journal = {Phys. Rev. D},
  volume = {7},
  issue = {10},
  pages = {2850--2862},
  numpages = {0},
  year = {1973},
  month = {May},
  publisher = {American Physical Society},
  doi = {10.1103/PhysRevD.7.2850},
  url = {https://link.aps.org/doi/10.1103/PhysRevD.7.2850}
}

@article{Hawking1974,
  author = {Hawking, S. W.},
  title = {{Black hole explosions?}},
  journal = {Nature},
  year = {1974},
  volume = {248},
  number = {5443},
  pages = {30--31},
  doi = {10.1038/248030a0},
  url = {https://doi.org/10.1038/248030a0}
}

@article{Hawking1975,
    author = "Hawking, S. W.",
    editor = "Gibbons, G. W. and Hawking, S. W.",
    title = "{Particle Creation by Black Holes}",
    doi = "10.1007/BF02345020",
    journal = "Commun. Math. Phys.",
    volume = "43",
    pages = "199--220",
    year = "1975",
    note = "[Erratum: Commun.Math.Phys. 46, 206 (1976)]"
}

@article{Davies1975,
doi = {10.1088/0305-4470/8/4/022},
url = {https://dx.doi.org/10.1088/0305-4470/8/4/022},
year = {1975},
month = {apr},
publisher = {},
volume = {8},
number = {4},
pages = {609},
author = {P C W Davies},
title = {{Scalar production in Schwarzschild and Rindler metrics}},
journal = {Journal of Physics A: Mathematical and General},
}

@book{Einstein100,
  author    = {DeWitt, Bryce S. },
  title     = {{General Relativity}: {An Einstein Centenary Survey}},
  isbn      = {978-0-521-29928-2},
  publisher = {Univ. Pr.},
  address   = {Cambridge, UK},
  year      = {1979}
}

@article{UnruhWald1984,
  title = {{What happens when an accelerating observer detects a Rindler particle}},
  author = {Unruh, William G. and Wald, Robert M.},
  journal = {Phys. Rev. D},
  volume = {29},
  issue = {6},
  pages = {1047--1056},
  numpages = {0},
  year = {1984},
  month = {Mar},
  publisher = {American Physical Society},
  doi = {10.1103/PhysRevD.29.1047},
  url = {https://link.aps.org/doi/10.1103/PhysRevD.29.1047}
}

@book{Birrell_Davies1982,
    author = "Birrell, N. D. and Davies, P. C. W.",
    title = "{Quantum Fields in Curved Space}",
    doi = "10.1017/CBO9780511622632",
    isbn = "978-0-511-62263-2, 978-0-521-27858-4",
    publisher = "Cambridge University Press",
    address = "Cambridge, UK",
    series = "Cambridge Monographs on Mathematical Physics",
    year = "1982"
}

@book{Wald1994,
    author = "Wald, Robert M.",
    title = "{Quantum Field Theory in Curved Space-Time and Black Hole Thermodynamics}",
    isbn = "978-0-226-87027-4",
    publisher = "University of Chicago Press",
    address = "Chicago, IL",
    series = "Chicago Lectures in Physics",
    year = "1994"
}

@book{Mukhanov2007,
  title = "{{Introduction to Quantum Effects in Gravity}}",
  author = {Mukhanov, Viatcheslav and Winitzki, Sergei},
  year = {2007},
  publisher = {Cambridge University Press},
  doi = {10.1017/CBO9780511809149}
}

@article{Colosi2009Rovelli,
doi = {10.1088/0264-9381/26/2/025002},
url = {https://dx.doi.org/10.1088/0264-9381/26/2/025002},
year = {2008},
month = {dec},
publisher = {},
volume = {26},
number = {2},
pages = {025002},
author = {Colosi, Daniele and Rovelli, Carlo},
title = {{What is a particle?}},
journal = {Classical and Quantum Gravity},
}

@article{Svidzinsky2021PRL,
  title = "{Unruh and Cherenkov Radiation from a Negative Frequency Perspective}",
  author = {Svidzinsky, Anatoly and Azizi, Arash and Ben-Benjamin, Jonathan S. and Scully, Marlan O. and Unruh, William},
  journal = {Phys. Rev. Lett.},
  volume = {126},
  number = {6},
  pages = {063603},
  year = {2021},
  publisher = {American Physical Society},
  doi = {10.1103/PhysRevLett.126.063603},
  url = {https://link.aps.org/doi/10.1103/PhysRevLett.126.063603}
}

@article{Svidzinsky2021PRR,
  title = "{Causality in quantum optics and entanglement of Minkowski vacuum}",
  author = {Svidzinsky, Anatoly and Azizi, Arash and Ben-Benjamin, Jonathan S. and Scully, Marlan O. and Unruh, William},
  journal = {Phys. Rev. Res.},
  volume = {3},
  number = {1},
  pages = {013202},
  year = {2021},
  publisher = {American Physical Society},
  doi = {10.1103/PhysRevResearch.3.013202},
  url = {https://link.aps.org/doi/10.1103/PhysRevResearch.3.013202}
}

@article{Barcelo2011LRR,
  author = {Barcel{\'o}, Carlos and Liberati, Stefano and Visser, Matt},
  title = "{{Analogue Gravity}}",
  journal = {Living Reviews in Relativity},
  volume = {14},
  number = {1},
  pages = {3},
  year = {2011},
  doi = {10.12942/lrr-2011-3},
  url = {https://doi.org/10.12942/lrr-2011-3},
  issn = {1433-8351}
}

@article{Barcelo:2005fc,
    author = "Barcelo, Carlos and Liberati, Stefano and Visser, Matt",
    title = "{Analogue gravity}",
    eprint = "gr-qc/0505065",
    archivePrefix = "arXiv",
    doi = "10.12942/lrr-2005-12",
    journal = "Living Rev. Rel.",
    volume = "8",
    pages = "12",
    year = "2005"
}

@article{Gooding:2020scc,
    author = "Gooding, Cisco and Biermann, Steffen and Erne, Sebastian and Louko, Jorma and Unruh, William G. and Schmiedmayer, Joerg and Weinfurtner, Silke",
    title = "{Interferometric Unruh detectors for Bose-Einstein condensates}",
    eprint = "2007.07160",
    archivePrefix = "arXiv",
    primaryClass = "gr-qc",
    doi = "10.1103/PhysRevLett.125.213603",
    journal = "Phys. Rev. Lett.",
    volume = "125",
    number = "21",
    pages = "213603",
    year = "2020"
}

@article{langen2013local,
  title={Local emergence of thermal correlations in an isolated quantum many-body system},
  author={Langen, Tim and Geiger, Remi and Kuhnert, Maximilian and Rauer, Bernhard and Schmiedmayer, Joerg},
  doi = "10.1038/nphys2739",
  journal={Nature Physics},
  volume={9},
  number={10},
  pages={640--643},
  year={2013},
  publisher={Nature Publishing Group}
}

@article{torres2017rotational,
  title={Rotational superradiant scattering in a vortex flow},
  author={Torres, Theo and Patrick, Sam and Coutant, Antonin and Richartz, Mauricio and Tedford, Edmund W and Weinfurtner, Silke},
  doi = "10.1038/nphys4151",
  journal={Nature Physics},
  volume={13},
  number={9},
  pages={833--836},
  year={2017},
  publisher={Nature Publishing Group}
}

@article{Martín-Martinez2011Berry,
  title = {{Using Berry's Phase to Detect the Unruh Effect at Lower Accelerations}},
  author = {Mart\'{\i}n-Mart\'{\i}nez, Eduardo and Fuentes, Ivette and Mann, Robert B.},
  journal = {Phys. Rev. Lett.},
  volume = {107},
  issue = {13},
  pages = {131301},
  numpages = {5},
  year = {2011},
  month = {Sep},
  publisher = {American Physical Society},
  doi = {10.1103/PhysRevLett.107.131301},
  url = {https://link.aps.org/doi/10.1103/PhysRevLett.107.131301}
}

@misc{Azizi2025WIT,
  author  = {Azizi, Arash},
  title   = {{Worldline-Induced Transparency}},
  year    = {2025},
  note    = {To appear}
}

@article{Azizi2025TPE,
  title = {{Two-photon emission from uniform acceleration: Unruh excitation, radiative decay, and field entanglement}},
  author = {Azizi, Arash},
  journal = {Phys. Rev. D},
  volume = {112},
  issue = {4},
  pages = {045008},
  numpages = {7},
  year = {2025},
  month = {Aug},
  publisher = {American Physical Society},
  doi = {10.1103/bck9-x7gt},
  url = {https://link.aps.org/doi/10.1103/bck9-x7gt}
}

@article{Azizi2025Dyson-Unruh,
  title = {{Time-ordering in the Dyson-Unruh problem: Accelerated observers and quantum fields}},
  author = {Azizi, Arash},
  journal = {Phys. Rev. D},
  volume = {112},
  issue = {2},
  pages = {025006},
  numpages = {16},
  year = {2025},
  month = {Jul},
  publisher = {American Physical Society},
  doi = {10.1103/cty8-mtt8},
  url = {https://link.aps.org/doi/10.1103/cty8-mtt8}
}

@article{Garay2000Sonic,
  title = {Sonic Analog of Gravitational Black Holes in Bose-Einstein Condensates},
  author = {Garay, L. J. and Anglin, J. R. and Cirac, J. I. and Zoller, P.},
  journal = {Phys. Rev. Lett.},
  volume = {85},
  issue = {22},
  pages = {4643--4647},
  numpages = {0},
  year = {2000},
  month = {Nov},
  publisher = {American Physical Society},
  doi = {10.1103/PhysRevLett.85.4643},
  url = {https://link.aps.org/doi/10.1103/PhysRevLett.85.4643}
}

@article{Retzker2008Unruh_BEC,
  title = {Methods for Detecting Acceleration Radiation in a Bose-Einstein Condensate},
  author = {Retzker, A. and Cirac, J. I. and Plenio, M. B. and Reznik, B.},
  journal = {Phys. Rev. Lett.},
  volume = {101},
  issue = {11},
  pages = {110402},
  numpages = {4},
  year = {2008},
  month = {Sep},
  publisher = {American Physical Society},
  doi = {10.1103/PhysRevLett.101.110402},
  url = {https://link.aps.org/doi/10.1103/PhysRevLett.101.110402}
}

@article{Carusotto2008Numerical_BEC,
doi = {10.1088/1367-2630/10/10/103001},
url = {https://doi.org/10.1088/1367-2630/10/10/103001},
year = {2008},
month = {oct},
publisher = {},
volume = {10},
number = {10},
pages = {103001},
author = {Carusotto, Iacopo and Fagnocchi, Serena and Recati, Alessio and Balbinot, Roberto and Fabbri, Alessandro},
title = {Numerical observation of Hawking radiation from acoustic black holes in atomic Bose–Einstein condensates},
journal = {New Journal of Physics},
}

@article{Lahav2010Sonic_BEC,
  title = {Realization of a Sonic Black Hole Analog in a Bose-Einstein Condensate},
  author = {Lahav, Oren and Itah, Amir and Blumkin, Alex and Gordon, Carmit and Rinott, Shahar and Zayats, Alona and Steinhauer, Jeff},
  journal = {Phys. Rev. Lett.},
  volume = {105},
  issue = {24},
  pages = {240401},
  numpages = {4},
  year = {2010},
  month = {Dec},
  publisher = {American Physical Society},
  doi = {10.1103/PhysRevLett.105.240401},
  url = {https://link.aps.org/doi/10.1103/PhysRevLett.105.240401}
}

@article{Jaskula2012Casimir_BEC,
  title = {Acoustic Analog to the Dynamical Casimir Effect in a Bose-Einstein Condensate},
  author = {Jaskula, J.-C. and Partridge, G. B. and Bonneau, M. and Lopes, R. and Ruaudel, J. and Boiron, D. and Westbrook, C. I.},
  journal = {Phys. Rev. Lett.},
  volume = {109},
  issue = {22},
  pages = {220401},
  numpages = {5},
  year = {2012},
  month = {Nov},
  publisher = {American Physical Society},
  doi = {10.1103/PhysRevLett.109.220401},
  url = {https://link.aps.org/doi/10.1103/PhysRevLett.109.220401}
}

@article{Eckel2018Expanding_Universe,
  title = {A Rapidly Expanding Bose-Einstein Condensate: An Expanding Universe in the Lab},
  author = {Eckel, S. and Kumar, A. and Jacobson, T. and Spielman, I. B. and Campbell, G. K.},
  journal = {Phys. Rev. X},
  volume = {8},
  issue = {2},
  pages = {021021},
  numpages = {13},
  year = {2018},
  month = {Apr},
  publisher = {American Physical Society},
  doi = {10.1103/PhysRevX.8.021021},
  url = {https://link.aps.org/doi/10.1103/PhysRevX.8.021021}
}
\end{document}